\documentclass[onecolumn,showkeys,preprintnumbers,amsmath,amssymb,floatfix]{revtex4}

\usepackage{graphicx}
\usepackage{dcolumn}
\usepackage{bm}
\usepackage{color}

\newcommand{\beq}{\begin{equation}}
\newcommand{\eeq}{\end{equation}}
\newcommand{\bey}{\begin{eqnarray}}
\newcommand{\eey}{\end{eqnarray}}

\begin{document}

\preprint{}

\title{Gravitational collapse of anisotropic stars}

 \author{Shyam Das}
  \email{dasshyam321@gmail.com}
\affiliation{Department of Physics, P. D. Women's College, Jalpaiguri 735101, West Bengal, India} 

\author{Bikash Chandra Paul}
 \email{bcpaul@associates.iucaa.in}
 \affiliation{Department of Physics, North Bengal University, Siliguri 734013, West Bengal, India}

\author{Ranjan Sharma}
\email{rsharma@associates.iucaa.in}
\affiliation{Department of Physics, Cooch Behar Panchanan Barma University, Cooch Behar, 736101, India}

\date{\today}

\begin{abstract}
We study the gravitational collapse of a spherically symmetric anisotropic relativistic star within Einstein’s theory of gravity making use of one of our recently developed collapsing stellar models [{\it Astrophys. Space Sci.} {\bf361} 99 (2016)]. The final state of continual gravitational collapse of a massive star under regular initial conditions is analyzed in terms of the formation of black holes. To study the evolution of an anisotropic star undergoing gravitational collapse, it is assumed that the dissipation process happens in the form of radial heat-flux. The interior space-time is described by static metric matched at the boundary with Vaidya metric that describes the exterior to the radiating star. The initial static configuration  is described by the relativistic solution obtained by Paul and Deb [{\it Astrophys. Space Sci.} {\bf354} 421 (2014)]. The impact of anisotropy on the dynamical gravitational collapse of a massive star is studied. The relativistic causal heat transport equation of the Maxwell-Cattaneo equation is utilized to show the dependence of anisotropy on the temperature profile of the collapsing system. 
\end{abstract}

\maketitle


\section{\label{sec1} Introduction} 

One of the most challenging issues in astrophysics and gravitation theory is that of dealing with the classical problem of gravitational collapse. Equilibrium gravitating structures like planets, stars, galaxies - all are considered to be formed through the process of gravitational collapse. The physics of gravitational collapse of stars has been studied extensively in astrophysics, in particular. 

A star maintains its equilibrium state as long as it produces outward thermal pressure through burning of its nuclear fuel; mainly hydrogen fusing into helium and later other heavier elements to counterbalance its inward gravitational pull. When the limited nuclear fuel supply exhausts completely, it can not halt its own gravitational pull and ultimately undergoes a contraction. This physical process is described as gravitational collapse and at the end it settles down to a new state of equilibrium. Given this situation, the effects of general relativity become the key factor in determining its final state.  Hence, it is necessary to study the dynamical gravitational collapse within the framework of Einstein's theory of gravity. In the 1960's, the discovery of the quasi-stellar radio sources, very high energy phenomena such as the gamma-ray bursts that release enormous amounts of energy in the form of gravitational radiation were explained as a result of the gravitational collapse of massive stars\cite{Hoyle}.

Depending upon the mass of a collapsing star, the final equilibrium state of gravitational collapse under the force of its own gravity at the end of its life cycle can be different in nature. The star will end as a white dwarf in which gravity is opposed by the electron degeneracy pressure if the mass of the collapsing core is less than $1.4$ solar mass which is the Chandrashekhar limit\cite{Chandra+1935}. It will end as a neutron star for which the mass is $\approx 0.7$ solar mass as calculated by Oppenheimer-Volkoff\cite{Oppen+Volkoff+1939}. In the later case, gravity is opposed by the neutron degeneracy pressure if the core has a mass greater than the Chandrashekhar limit and less than about $3-5$ times the mass of the sun. For a sufficiently large mass (of the order of about ten solar masses or more) the thermal energy is radiated away much quickly and no known form of internal pressure can balance the inward pull due to gravity. Without settling into any equilibrium state, the end stage of such a catastrophic collapse results in the formation of space-time singularity which may be covered by an event horizon - a black hole or a naked singularity, depending upon the initial conditions of the collapsing configuration. According to singularity theorems, a sufficiently massive collapsing object will undergo continual gravitational collapse, resulting in the formation of a gravitational singularity. Event horizon telescope provides an excellent observational evidence in support of astrophysical black holes. 

In relativistic astrophysics, one of the central questions is to investigate the final fate of massive stars. By choosing the appropriate geometry of interior and exterior regions, investigators are trying to understand the underlying physics. The initial step to study the gravitational collapse was taken by Oppenheimer and Synder\cite{Oppen+Snyder+1939} and Datt\cite{Datt+1938}(OSD). In the OSD model, under the assumptions of homogeneity, sphericity and dust equation of state, they have shown that the end state of gravitational collapse leads to the formation of a black hole. Cosmic censorship conjecture by Penrose\cite{Penrose} ensures that the gravitational collapse of any realistic matter distribution ends with the formation of a black hole similar to the OSD model. There is no theoretical or mathematical proof available to this conjecture though. By relaxing homogeneity condition, Datt-Tolman-Bondi have found a solution that describes the collapse of a dust sphere with non-uniform initial density. They have shown that depending on the initial conditions, the collapse ends with the formation of two kinds of singularities - black hole and a naked singularity. The construction of more realistic gravitational collapse models was possible after Vaidya\cite{Vaidya+1951} discovered a solution describing the exterior gravitational field of a stellar body with outgoing radiation and Santos\cite{Santos+1985} developed the junction conditions at the boundary joining the interior space-time of the collapsing object to the Vaidya exterior metric. Self-similar gravitational collapse of a perfect fluid with a linear equation of state was considered by Ori and Piran\cite{Ori+Piran}. The collapse leads to a curvature singularity which can be either naked or hidden. Such a scenario was also analyzed by Joshi and Dwivedi\cite{Joshi+Dwi}.  Considering the collapsing fluid to be imperfect in nature, collapse properties have been analyzed by Szekeres and Iyer\cite{S+I}. Recently, collapse of a fluid with tangential pressure has been studied\cite{Magli,Singh+W}. By introducing the cosmological constant, the effects of positive as well as negative cosmological constant on gravitational collapse have been studied by Markovic and Shapiro\cite{Markovic+S} and Lake\cite{Lake}. The quasi-spherical collapse has been studied by Debnath {\em et al}\cite{Debnath}. Sharif and Ahmad\cite{Sharif+A1,Sharif+A2} have studied the gravitational collapse of a perfect fluid in higher dimensional space-time. The gravitational collapse of a scalar field has been studied by many investigators\cite{Christodoulou,Giambo+2005,Bhattacharjee+2010}. In the framework of alternative theories of gravity such as $f(R)$ gravity, it has been reported that both black hole and a naked singularity are possible so far as the final outcome of a gravitational collapsing star is concerned\cite{Ziaie+2010,Bedjaoui+2010,Ohashi+2011}.

It is noteworthy that gravitational collapse is a highly dissipating process. To understand the dynamics of dissipation during the collapse process, impacts of factors like bulk viscosity, anisotropic pressure, shear, electromagnetic field etc. have been introduced. Realistic stellar models have been developed where impacts of many of these factors have been taken up and studied\cite{Bonnor+et al+1989,Govender+et+al+2018,Oliveira+et al+1988,Oliveira+Kolassis+Santos+1988,Herrera+Ospino+Prisco+2008,Herrera+Prisco+Pastora+Santos+1998,Herrera+Santos+1997,Herrera+Prisco+Martin+Ospino+Santos+Troconis+2004,Herrera+Barrreto+2011,Herrera+et al+1997,Herrera+et al+2002,Herrera+et al+2011,Prisco+et al+2007,Prisco+et al+1997,Martinez+1996, Pinheiro+Chan,Chan+2000,Chan+2003,Chan+1994,Tikekar+Patel+1992,Maharaj+Govender+2000,Mena+et al+2004,Thiru+Mharaj+2009,Govinder+et al+1998,Dirk+et al+2000,Ivanov,Sarwe+Tikekar+2010,Sharma+Tikekar+2012,Sharma2+Tikekar+2012,Sharma+et al+2015}. Of particular interest is the effects of pressure anisotropy, which is the difference of pressures generated in a system between radial and tangential directions. In relativistic astrophysics, the consideration of anisotropic stress in a collapsing system originates from various factors. Ruderman\cite{Ruderman+1972} and Canuto\cite{Canuto+1974} have shown that in the high-density regime of the compact stellar interior, anisotropy may develop. In astrophysical objects, the existence of a solid core or type $3$A super fluid\cite{Kippen+Weigert+1990}, strong magnetic field\cite{Weber+1999}, slow rotation\cite{Herrera+Santos+1995}, phase transition\cite{Sokolov+1980}, pion condensation\cite{Sawyer+1972}, strong electromagnetic field\cite{Usov+2004} etc. can generate anisotropy.  It has also been shown that a mixture of perfect and a null fluid effectively provides an anisotropic fluid model\cite{Letelier+1980}. Similarly, it has been pointed out by Ivanov\cite{Ivanov+2010} that shear, charge, etc. on self-bound systems can be absorbed if one considers the system to be anisotropic, in general. The shearing motion of the fluid can be considered as one of the reasons for the presence of anisotropy in a self-gravitating body\cite{Chan+1997, Chan+1998, Chan+1993}. Bowers and Liang\cite{Bowers+Liang+1974} have extensively explored the underlying causes of pressure anisotropy in the stellar interior and analyzed the effects of anisotropic stress on the physical behaviour like radial pressure, critical mass and maximum surface redshift of a star undergoing gravitational collapse. Dev and Gleiser\cite{Dev+Gleiser+2002} have shown that critical mass and surface redshifts increase with anisotropy. Herrera and co-workers\cite{Herrera+et al+2008} have analyzed the impact of anisotropic stresses on gravitational collapse of fluid distributions. The analysis of Dev and Gleiser\cite{Dev+Gleiser+2003} and Chan\cite{Chan+1993} show that anisotropy provides a more stable configuration. Sharma and Das\cite{Sharma+Das+2013} have developed a model for a self-gravitating spherically symmetric relativistic star which begins to collapse from an initially static configuration by dissipating energy in the form of radial heat flow. The model provides a simple technique to examine the impact of pressure anisotropy on collapse. Reddy {\em et al}\cite{Reddy+et al+2015} have analyzed impacts of anisotropic pressure in a shearing condition onto the collapse, which starts collapsing from an initial static configuration described by Bowers and Liang\cite{Bowers+Liang+1974} static model. Govender {\it et al}\cite{Goven+et al+2016} have constructed a model of a collapsing star by incorporating dynamical effects into the Pant and Sah model. 

The aim of the current investigation is to analyze the effects of anisotropy on the collapse for a large variety of masses. Earlier, Das {\it et al}\cite{Das+et al+2016} have developed a model to study the effects of pressure anisotropy on the evolution of a collapsing star which begins its collapse from an initial static configuration described by Paul and Deb\cite{{pd+et al+2014}}. In this model, we extend our previous work to analyze the impact of anisotropy on the evolution of massive collapsing stars and temperature evolution during the collapse. It will be interesting to analyze how the different model parameters can affect the collapse of massive as well as lower masses stars. In our study, we shall investigate the problem of gravitational collapse for a spherically symmetric collapsing anisotropic star in the presence of dissipation.  

The paper is organized as follows. We first outline the model developed previously by Das {\it et al}\cite{Das+et al+2016} for an anisotropic matter distribution collapsing under the influence of its own gravity and dissipating energy in the form of radial heat flux. In Sec.~\ref{sec2}, the Einstein field equations are laid down. The exterior space-time of the collapsing matter is appropriately described by the Vaidya metric and in Sec.~\ref{sec3}, we outline the main results of the junction conditions joining smoothly the interior ($r < r_{\Sigma}$) and the exterior ($r > r_{\Sigma}$) space-times across the boundary surface $\Sigma$ separating the two regions.  Making use of the junction conditions and a static seed solution, we develop collapsing model in Sec.~\ref{sec4}. In Sec.~\ref{sec5}, we  analyze the effects of pressure anisotropy. We discuss the thermal behaviour of the collapsing star in Sec.~\ref{sec6}. Finally, we conclude by discussing our results in Sec.~\ref{sec7}.
  
\section{\label{sec2}Einstein  field equations}
We write the interior spacetime of a gravitationally collapsing star in isotropic coordinates as
\begin{equation}
ds_{-}^2 = -A^2(r,t)dt^2 + B^2(r,t)\left[dr^2+r^2(d\theta^2 + \sin^2\theta d\phi^2)\right],\label{intm1}
\end{equation}
where $A(r,t)$ and $B(r,t)$ are yet to be specified.  We assume the matter composition of the collapsing star to be an imperfect fluid with anisotropic stresses undergoing dissipation in the form of radial heat flow. Accordingly, energy-momentum tensor of the collapsing fluid is written in the form
\begin{equation}
T_{ij} = (\rho + p_t)u_i u_j + p_t g_{ij} + (p_r-p_t)X_i X_j + q_i u_j + q_j u_i,\label{emt1}
\end{equation}
where $\rho$ is the energy density of the fluid, $p_r$ is the radial pressure, $p_t$ is the tangential pressure, $q^i=q\;\delta_1^i$ is the radial heat flux vector, $X^i$ is an unit $4$-vector along the radial direction and $u^i$ is the $4$-velocity of the fluid. The quantities obey the following relations: $u^i q_i = 0$, $X_i X^i = 1$, $X_i u^i = 0$.

The Einstein field equations for the line element (\ref{intm1}) are obtained as
\begin{equation}
\kappa^2 \rho =-\frac{1}{B^2}\left(\frac{2B''}{B}+\frac{4B'}{rB}-\frac{B'^2}{B^2}\right)+\frac{3\dot{B}^2}{ A^2 B^2}, \label{Eq1}
\end{equation}
\[
\kappa^2 (p_r-\zeta \Theta) =\frac{1}{B^2}\left(\frac{B'^2}{B^2}+\frac{2B'}{rB}+\frac{2A'B'}{AB}+\frac{2A'}{rA}\right)   
\]
\begin{equation}
\; \; \;  \; \; \; \; \; \; \; \; \; \; \; +\frac{1}{A^2}\left(-\frac{2\ddot{B}}{B}-\frac{\dot{B}^2}{B^2}+\frac{2\dot{A}\dot{B}}{AB}\right), \label{Eq2}
\end{equation}
\[
\kappa^2 (p_t-\zeta \Theta) = -\frac{2\ddot{B}}{A^2 B}+\frac{2\dot{A}\dot{B}}{A^3 B}-\frac{\dot{B}^2}{A^2 B^2}  
+\frac{A'}{r A B^2}
\]
\begin{equation}
   \hspace{0.8 cm } \;   +\frac{B'}{r B^3}+\frac{A''}{A B^2}-\frac{B'^2}{B^4}+\frac{B''}{B^3}, \label{Eq3}
\end{equation}

\begin{equation}
\kappa^2 q = -\frac{2}{A B^2}\left(-\frac{\dot{B}'}{B}+\frac{\dot{B}B'}{B^2}+\frac{A'\dot{B}}{AB}\right), \label{Eq4}
\end{equation}
where, we  set $c=1$ and  $8\pi G = \kappa^2$. Note that a prime ($'$) and an overhead dot ($.$) denote differentiation with respect to $r$ and $t$, respectively.


\section{\label{sec3}Exterior space-time and junction conditions}

The space-time exterior to a radiating star is appropriately described by the Vaidya\cite{Vaidya+1951} metric
\begin{equation}
ds_{+}^2 = -\left(1-\frac{2m(v)}{{\sf r}}\right)dv^2 - 2dv d{\sf r} + {\sf r}^2[d \theta^2 + \sin^2 \theta d\phi^2], \label{Vm}
\end{equation}
where the mass function $m(v)$ is a function of retarded time $v$. The junction conditions for smooth joining of  the interior (\ref{intm1}) and exterior (\ref{Vm}) spacetimes are obtained by assuming a time-like $3$-surface $\Sigma$ that separates the interior and the exterior manifolds\cite{Santos+1985}. Continuity of metric functions ($(ds_{-}^2)_{\Sigma} = (ds_{+}^2)_{\Sigma} = ds_{\Sigma}^2$) and extrinsic curvatures ($K_{ij}^{-} = K_{ij}^{+}$) across the surface $\Sigma$, then yield the following matching conditions for the interior ($r \leq r_\Sigma$) and  for the exterior ($ r \geq r_\Sigma$) space-times at the boundary:
\begin{equation} 
m(v) = \left(\frac{r^3 B \dot{B}^2}{2 A^2}-r^2 B' -\frac{r^3 B'^2}{2 B}\right)_{\Sigma}, \label{mm}
\end{equation}
\begin{equation}
(p_r)_\Sigma =(q B)_\Sigma.\label{eqj}
\end{equation}

Consequently, using Eq.~(\ref{mm}), one can determine the mass of the evolving star at any instant $t$ within a radius $r < r_{\Sigma}$ as
\begin{equation}
m(r,t) \stackrel{\Sigma}{=} \left(\frac{r^3 B \dot{B}^2}{2 A^2}-r^2 B' -\frac{r^3 B'^2}{2 B}\right).\label{inmass}
\end{equation}

In the next section, a method to solve the system of equations is provided which will help us to study the behaviour of the collapsing star. The technique was earlier provided by Das {\it et al}\cite{Das+et al+2016}.

\section{\label{sec4} A Toy model}
By combining Eqs.~(\ref{Eq2}) and (\ref{Eq3}), we obtain the anisotropy which is given by
\begin{equation}
\Delta(r,t)=\frac{1}{B^2}\left[\frac{B''}{B}+\frac{A''}{A}-\frac{2 B'^2}{B^2}-\frac{B'}{r B}-\frac{2 A' B'}{A B}-\frac{A'}{r A}\right],\label{ani}
\end{equation}
where, $\Delta(r,t) = \kappa^2 (p_t-p)$ is the measure of anisotropy of the collapsing matter. Now, to make eq.~(\ref{ani}) tractable, we assume that the metric potentials $A(r,t)$, $B(r,t)$ are separable in their variables and accordingly we write
\begin{eqnarray}
A(r,t) &=& A_0(r), \label{metA}\\
B(r,t) &=& B_0(r) h(t). \label{metB}
\end{eqnarray}
We further define
\begin{equation}
\Delta(r,t) = \frac{\Delta_s(r)}{h^2(t)},\label{ani2}
\end{equation}
where $\Delta_s$ is the anisotropy for a static configuration of a relativistic star.
Using eq.~(\ref{ani}) we define $\Delta_s$ which is given by
\begin{equation}
\Delta_s(r) = \frac{1}{B_0^2}\left[\frac{B_0''}{B_0}+\frac{A_0''}{A_0}-\frac{2 B_0'^2}{B_0^2}-\frac{2 A_0' B_0'}{A_0 B_0}-\frac{A_0'}{r A_0}-\frac{B_0'}{B_0 r}\right].\label{ani3}
\end{equation}
Therefore, a physically reasonable static anisotropic stellar solution ($A_0(r), B_0(r)$) satisfying Eq.~(\ref{ani3}) is useful to study stellar collapse. The Einstein field eqs. (\ref{Eq1})-(\ref{Eq4}) can be written as
\begin{eqnarray}
\rho &=& \frac{\rho_0}{h^2}+\frac{3 \dot{h}}{\kappa^2 h^2 A_0^2},\label{fEq1}\\
p_r &=& \frac{p_0}{h^2}-\frac{1}{\kappa^2 A^2_0}\left(\frac{2\ddot{h}}{h}+\frac{\dot{h}^2}{h^2}\right)
,\label{fEq2}\\
p_t &=& \frac{p_{t0}}{h^2}-\frac{1}{\kappa^2  A^2_0}\left(\frac{2\ddot{h}}{h}+\frac{\dot{h}^2}{h^2}\right)
,\label{fEq3}\\
q &=& -\frac{2A'_0\dot{h}}{\kappa^2 A^2_0 B^2_0 h^3},\label{fEq4}
\end{eqnarray}
where, 
\begin{eqnarray}
\rho_0 &=& -\frac{1}{\kappa^2 B_0^2}\left[\frac{2B_0''}{B_0}+\frac{4B_0'}{rB_0}-\frac{{B_0'}^2}{B_0^2}\right],\label{sden}\\
p_{r0} &=& \frac{1}{\kappa^2 B_0^2}\left[\frac{B_0'^2}{B_0^2}+\frac{2B_0'}{rB_0}+\frac{2A'_0 B'_0}{A_0 B_0}+\frac{2A'_0}{rA_0}\right],\label{spr}\\
p_{t0} &=& \frac{1}{\kappa^2 B^2_0 r^4} \left[\frac{B_0''}{B_0}+\frac{B_0'}{r B_0}+\frac{A_0''}{A_0}+\frac{A_0'}{r A_0}-\frac{B_0'^2}{B_0^2}\right].\label{spt}
\end{eqnarray}
In our model, we assume that the collapse begins from an initial anisotropic distribution of matter in a geometry obtained by Paul and Deb\cite{pd+et al+2014}, which is given by
\begin{eqnarray}
A_0(r) &=& a\left(\frac{1-k \alpha+n\frac{r^2}{R^2}}{1+k\alpha}\right),\label{pd1}\\
B_0(r) &=& \frac{(1+k \alpha)^2}{1+\frac{r^2}{R^2}},\label{pd2}
\end{eqnarray}
with $a$ a constant and 
\begin{equation}
\alpha(r) = \sqrt{\frac{1+\frac{r^2}{R^2}}{1+\frac{\lambda r^2}{R^2}}},\label{pd3}
\end{equation}
which contain four parameters $\lambda$, $k$, $n$ and $R$. Here the anisotropy of a compact object depends on the non zero values of the  parameter $n$ as $n=0$ leads to isotropy which will be described in the next section. It is interesting to note that $p_r > p_t$ corresponds to $n<0$ and $p_r < p_t$ corresponds to $n>0$. For a given mass and radius, the parameters can be fixed by using appropriate boundary conditions. Details of the numerical procedure to evaluate the values of constraints and physical viability of the model are discussed in  {\it Ref.}\cite{pd+et al+2014}. For appropriate choice of the model parameters,  it is shown that the relativistic solution  satisfies the causality condition as well as strong energy condition accommodating realistic stars. Thus in a compact star, collapse may be  studied for a given star which begins its collapse from a state with reasonable initial conditions when equilibrium is lost. 

The dynamical quantities of the  initial configuration of a star in this case are given below:
\begin{eqnarray}
\rho_0 &=& \frac{12(1+ k \alpha^5 \lambda)}{\kappa^2 R^2(1+k \alpha)^5},\label{sden0}\\
p_{r0} &=&\frac{2(2 x^2(k^2 \alpha^6 \lambda-1)+n(2-\alpha^2 z))}{x R^2  (1+k\alpha)^5(n(\alpha^2-1)+(1-k\alpha)x)},\label{sp0}\\
p_{t0} &=& \frac{x(4 x^2(k^2 \alpha^6 \lambda-1)+n y)}{R^4 v(\alpha^2-1)^2(n(\alpha^2-1)+(1-k \alpha)x)},\label{spt0}\\
\Delta_s &=& \frac{n \alpha(1-\alpha^2)(-8\alpha(\lambda-1)+k(-3+u ))}{ v (\alpha^2\lambda-1)(n(\alpha^2-1)-(k\alpha-1) x)   },\label{sani0}
\end{eqnarray}
where, we have defined
\begin{eqnarray}
x &=& (1-\alpha^2 \lambda),\label{c1}\\
y &=& (4 + \alpha (4 \alpha (\alpha^2 -1) + k (3 + \alpha^4) - 4 \alpha (1 + 2 k \alpha+ \alpha^2) \lambda \nonumber\\
&& + \alpha^3 (4 + k \alpha (9 - 8 \alpha^2 + 3 \alpha^4) \lambda^2))) \label{c2},\\
z &=& (-2 +6 \lambda+\alpha(-2\alpha(-1+\lambda+\lambda^2)\nonumber\\
&&-k(1+\lambda+\alpha^2(1+\lambda^2(5 \alpha^2-3)-\lambda(2+3 \alpha^2))))),\\
u& = &   \alpha^2(-1+\lambda(10+3\alpha^2(2+(-5+\alpha^2)\lambda))), \\
v &=&  R^2(1+k \alpha)^5. \label{c3}
\end{eqnarray}

Using eq.~(\ref{eqj}) together with the condition that pressure of the initial static configuration must vanish at the surface, {\it  i.e., }$p_0(r = r_\Sigma) = 0$, we obtain 
\begin{equation}
2h\ddot{h}+\dot{h}^2-2 \gamma \dot{h} =0,\label{sur}
\end{equation}
where,
\begin{equation}
\gamma = \left[\frac{A_0'}{B_0}\right]_\Sigma.\label{surcon}
\end{equation}
Note that for a given initial static configuration $\gamma$ is a constant. Solution of eq.~(\ref{sur}) governs the overall temporal behaviour of the collapsing star. A similar approach was adopted by  Govender  {\em et al}\cite{Govender+et+al+2018} for solving junction conditions so as to study the impact of anisotropy on the time of formation of the horizon and the evolution of the temperature. 

Following Bonnor\cite{Bonnor+et al+1989}, we express the second order differential equation into a first order differential equation
\begin{equation}
\dot{h} = -\frac{2\gamma}{\sqrt{h}}(1-\sqrt{h}),\label{surf}
\end{equation}
which admits a solution 
\begin{equation}
t = \frac{1}{\gamma}\left[\frac{h}{2}+\sqrt{h}+\ln(1-\sqrt{h})\right].\label{eq}
\end{equation}
Obviously, at $t \rightarrow - \infty$, $h = 1$ and $\dot{h} =0$ when $h =1$. Thus, in this construction, the collapse begins in the remote past $t \rightarrow - \infty$ when $h = 1$ and it continues till the black hole is formed. 

Using eq.~(\ref{inmass}), we write the mass of the evolving star at any instant $t$ within a radius $r$ as
\begin{equation}
m(r,t) \stackrel{\Sigma}{=} \left[m_0 h(t)+\frac{2 \gamma^2 r^3 B_0^3(1-\sqrt{h})^2}{A_0
^2}\right],\label{mass}
\end{equation}
where,
\begin{equation}
m_0 \stackrel{\Sigma}{=} -\frac{r^2 B_0'}{2}\left(2+\frac{B_0' r}{B_0}\right),\label{masstc}
\end{equation}
defined to be the mass of the initial static star. We assume that at  $t= - \infty$ (i.e., when $h(t) = 1$), the collapsing star had an initial mass $m_0$ within a boundary surface $r_0$. The star continues to collapse until it shrink to a radius approaching the horizon $[r h (t_{bh})]_\Sigma =r_0 h(t_{bh}) = 2m(v)$, where $t= t_{bh}$ represent the time of formation of the black hole. The condition yields
\begin{equation}
h(t)=\left[\frac{\frac{2 r \gamma B_0}{A_0}}{\frac{2 r \gamma B_0}{A_0}+\sqrt{1-\frac{2 m_0}{r B_0}}}\right]^2_\Sigma,\label{hts}
\end{equation}
which attains the value $h_{bh}$ at a time
\begin{equation}
t_{bh} = \frac{1}{\gamma}\left[\frac{h_{bh}}{2}+\sqrt{h_{bh}}+\ln(1-\sqrt{h_{bh}})\right],\label{tbh}
\end{equation}
where $h(t_{bh}) = h_{bh}$.

\section{\label{sec5}Physical analysis}
In this section, the effects of anisotropy onto the collapse of compact objects for different initial masses and radii with different compactness factor are studied. We note the following:

{\bf Case -I}: Consider a compact object with the initial configuration of mass $m_0 = 3~M_{\odot}$ and radius $r_0 = 10\times1.475~$km, with compactness factor $u=0.3$. It is found that in the stellar model, once the star departs from equilibrium configuration, it starts collapsing until it becomes a  black hole. For different values of the anisotropic parameter $n$ the final of a massive star {\it i.e.}, its black hole mass ($m_{bh}$)  and the corresponding horizon radius ($r_{bh}$) are determined. 
The time of formation of a black hole ($t_{bh}$) in addition of $m_{bh}$  and $r_{bh}$ for stars having different  anisotropies $n$ and $k$ for the given mass is tabulated in Table~\ref{tab1}. The value of $a$ turns out to be $1$ for all values of $n$. 
Though the variation of $t_{bh}$ is small in this configuration, we note that, if $p_r > p_t$,  {\it i.e.}, $n < 0$, the horizon is formed at a later stage in a star having more negative $n$ value.

{\bf Case -II}: Consider a compact object with the configuration of mass $m_0 = 5~M_{\odot}$ and radius $r_0 = 2159~$km\cite{Bonnor+et al+1989}. The star considered here has the compactness factor $u=0.00341$. It has been shown that, once the star departs from equilibrium, it starts collapsing until it becomes a  black hole. For different values of the anisotropic parameter $n$, the values of the model parameters including the black hole mass ($m_{bh}$) and its corresponding horizon radius ($r_{bh}$) are determined, which are tabulated in Table~\ref{tab2}. The time of formation of black hole ($t_{bh}$) for  stars having different  anisotropies $n$ for a given mass are tabulated in Table~\ref{tab2}. The value of $a$ turns out to be $1$ for all values of $n$. Though the variation of $t_{bh}$ is very small in this configuration, we note that, when the tangential pressure of the initial configuration is greater than the radial pressure ($p_t > p_r$), {\it  i.e., } $n > 0$, the horizon is found to form at an earlier epoch compared to a less anisotropic star. However, if $p_r > p_t$,  {\it i.e.}, $n < 0$, the horizon is formed at a later stage in a star having more negative $n$ values.

{\bf Case -III}: Consider a compact object with the configuration of mass $m_0=6~M_{\odot}$ and radius $r_0=25.742~$km\cite{Woosley+Phillips+1988} which has the compactness factor $u=0.343$. With the variation of the anisotropic parameter $n$, the values of the model parameters including the mass ($m_{bh}$) of the black hole and  the corresponding horizon radius ($r_{bh}$) are determined, which are presented  in Table~\ref{tab3}. The time of formation of black hole ($t_{bh}$) for  stars having different  anisotropies $n$ for a given mass are also tabulated in Table~\ref{tab3}. Here also for $p_t > p$, {\it  i.e., } $n > 0$, the horizon is found to be formed at an earlier epoch compared to a less anisotropic star. However, if $p_r > p_t$,  {\it i.e.}, $n < 0$, the horizon will be formed at a later stage in a star having more negative $n$ values.

{\bf Case -IV}: Consider a star having mass $m_0=10~M_{\odot}$, radius $r_0=40\times1.475~km$, $k=0.2$, compactness factor $u=0.25$.  The values of the model parameters namely, mass of the black hole ($m_{bh}$), corresponding black hole radius ($r_{bh}$) and time of black hole formation ($t_{bh}$) for different anisotropic parameter $n$, are displayed in Table~\ref{tab4}. It is noted that as the anisotropy increases the formation time of black hole from infinitely past decreases, thereafter it increases for negative $n$ values. It is also noted that stars with $m=10 M_{\odot}$ having different physical radius $2 r_0$, the black hole formation occurs at a late stage for stars with bigger radius with lower black hole mass to be the end product. In this case, the values of the model parameters namely, mass of the black hole ($m_{bh}$), corresponding black hole radius ($r_{bh}$) and time of black hole formation ($t_{bh}$) for different anisotropic parameter $n$, are displayed in Table~\ref{tab5}. 

{\bf Case -V}: Consider a massive compact star with the configuration of mass $m_0 = 20~M_{\odot}$ and radius $r_0 = 60\times1.475~km$. The star considered here has the compactness factor $u=0.333$. For different values of the anisotropic parameter $n$, the values of the model parameters including the mass ($m_{bh}$) of the black hole and the corresponding horizon radius ($r_{bh}$) are determined, which are presented in Table~\ref{tab6}. We note that, for $n < 0$, the horizon is formed at a later stage in a star as compared to its isotropic configuration ($n=0$). In this case, the black hole formation delayed for a star having more negative $n$ values.

{\bf Case -VI}: Consider the gravitational collapse of stars with higher initial static mass. For this purpose we consider star having initial mass $m_0=60~M_{\odot}$, physical radius $r_0 =600 \times1.475$ km., $k=0.1$ and having compactness factor $u=0.1$. In  Table~\ref{tab7}    we have complied all the parameters related to describe the collapsing system. We note that as the anisotropic parameter $n$ is decreased the mass of black hole ($m_{bh}$) and corresponding black hole radius ($r_{bh}$)  both decreases. However, the formation time of black hole  ($t_{bh}$) increases.
 
\begin{table}
\caption{\label{tab1} Dynamical variables for different anisotropic star. ( $m_0 = 3~M_{\odot}$, $r_0 = 10\times1.475~km$.)} 
\begin{tabular}{cccccccc}
 $n$ & $k$ &$\lambda$ & $R$ & $h_{bh}$ & $t_{bh}$ &  $r_{bh}$ & $m_{bh}~(M_{\odot})$\\ \hline
 0.5 & 0.2 &54.28 &  135.22 & 0.0722  & -1.582 & 1.410 & 0.4780 \\
         & 0.4 &835.5 &  197.24 & 0.2015  & -3.616 & 3.512 & 1.1654\\ \hline
   0 & 0.2 &123.19 &  194.08 & 0.0667  & -1.467 & 1.302 & 0.4413 \\
         & 0.4 &2106.07 & 276.84 & 0.1744  & -3.566 & 3.402 & 1.1534\\ \hline
-1.0 &0.2& 261.51 &  276.42 & 0.0650  & -1.408 & 1.240 & 0.4230 \\
         & 0.4 &3961.39 & 377.76 & 0.1714  & -3.512 & 3.344 & 1.1335 \\ \hline
-2.0 &0.2& 399.93 &  339.34 & 0.0630  & -1.389 & 1.229 & 0.4169 \\
     & 0.4 &5819.79 & 457.02 & 0.1703  & -3.491 & 3.322 & 1.1261\\ \hline
-3.0 &0.2& 538.39 &  392.29 & 0.0620  & -1.379 & 1.221 & 0.4138 \\
     & 0.4 &7679.29 & 524.47 & 0.1697  & -3.481 & 3.311 & 1.1223\\ \hline 
-5.0 &0.2& 815.34 &  481.01 & 0.0621  & -1.370 & 1.212 & 0.4108\\ 
     & 0.4 &11398.70 & 638.34 & 0.1691  & -3.470 & 3.299 & 1.1184\\ \hline
\end{tabular}
\end{table}

\begin{table}
\caption{\label{tab2} Dynamical parameters for different anisotropic star. ( $m_0 = 5~M_{\odot}$, $r_0 = 2159~km$ and $k= 0.01$)} 
\begin{tabular}{ccccccccc}
$n$ & $\lambda$ & $R$ &  $t_{bh}$ &  $r_{bh}$ & $m_{bh}~(M_{\odot})$ \\ \hline
  1.00  & 945869          & 367763          & -0.06733 & 0.100046 & 0.034056 \\
 0.90 & $4.151\times 10^7$ & $2.412\times 10^6$ & -0.06354 & 0.094814 & 0.032140 \\
0.80 & $8.129\times 10^7$ & $3.374\times 10^6$ & -0.06349 & 0.094746 & 0.032117 \\ 
0.70 & $1.214\times 10^8$ & $4.125\times 10^6$ & -0.06348 & 0.094732 & 0.032112\\
0.50 & $2.018\times 10^8$ & $5.317\times 10^6$ & -0.06347 & 0.094710 & 0.032105 \\
-0.10 & $4.429\times 10^8$ & $7.870\times 10^6$ & -0.06346 & 0.094699 & 0.032101 \\
-0.50 & $6.036\times 10^8$ & $9.195\times 10^6$ & -0.06345  & 0.094690 & 0.032098 \\
-1.00 & $8.046\times 10^8$ & $1.061\times 10^7$ & -0.06345 & 0.094687 & 0.032097 \\
-1.99 & $1.202\times 10^9$ & $1.297\times 10^7$  & -0.06345 & 0.094686 & 0.032097 \\
-2.99 & $1.604\times 10^9$ & $1.499\times 10^7$  & -0.06344 & 0.094685 & 0.032096 \\
-3.99 & $2.006\times 10^9$ & $1.676\times 10^7$  & -0.06302 & 0.094543 & 0.031870 \\ \hline
\end{tabular}
\end{table}

\begin{table}
\caption{\label{tab3} Dynamical parameters for different anisotropic star. ( $m_0 = 6~M_{\odot}$, $r_0 = 25.742~km$ and $k= 0.2$.)} 
\begin{tabular}{ccccccccc}
$n$ & $\lambda$ & $a$ & $R $ & $h_{bh}$ & $t_{bh}$ &  $r_{bh}$ & $m_{bh}(M_{\odot})$\\ \hline
0.7 & 9.81 & 1.0074 & 178.04 & 0.04246 & -1.8858 & 1.5009 & 0.5087 \\
0.4 &  29.20 & 1.0069 & 262.25 & 0.03844 & -1.7121 & 1.3579 & 0.4603 \\ 
0.1 & 48.74 & 1.0068 & 324.98 & 0.03696 & -1.6495 & 1.3066 & 0.4429 \\
-0.4 & 81.58 & 1.0067 & 409.04 & 0.03589 & -1.6036 & 1.2691 & 0.4301 \\
-0.7 & 101.27 & 1.0067 & 451.95 & 0.03555 & -1.5890 & 1.2571 & 0.4261 \\
-1.5 & 153.80 & 1.0066 & 550.23 & 0.03505 & -1.5670 & 1.2391 & 0.4200 \\
-2.5 & 219.49 & 1.0066 & 652.58 & 0.03474 &  -1.5537 & 1.2282 & 0.4163 \\
-5.5 & 416.60 & 1.0065 & 891.75 & 0.03438 & -1.5382 & 1.2156 & 0.4208 \\
-8.5 & 613.73 & 1.0065 & 1079.15 & 0.03425 & -1.5325 & 1.2109 & 0.4105 \\
-10.0 & 712.29 & 1.0065 & 1161.57 & 0.03421 & -1.5307 & 1.2096 & 0.4100 \\ 
-50.0 & 3340.74 & 1.00655& 2504.76&0.034021 & -1.5224 &1.2027 &0.40771\\ \hline
\end{tabular}
\end{table}

\begin{table}
\caption{\label{tab4} Dynamical variables for different anisotropic star. ( $m_0 =10 ~M_{\odot}$, $r_0 =40\times1.475~km$ and $k = 0.2$.)} 
\begin{tabular}{cccccccc}
$n$  & $\lambda$ & $a$ & $R $ & $h_{bh}$ & $t_{bh}$ &  $r_{bh}$ & $m_{bh}~(M_{\odot})$\\ \hline
0.80 &  67.32 & 1.0040 & 436.90 & 0.09911 & -7.359 & 7.40 & 2.51 \\ 
0.50 & 182.63 & 1.0044 & 671.53 & 0.08522 & -6.389 & 6.36 & 2.16 \\
0.30 & 260.49 & 1.0044 & 790.76 & 0.08228& -6.181 & 6.14& 2.08 \\
0.10 & 338.57 & 1.0044 & 894.37 & 0.08059& -6.063 & 6.02 & 2.04\\
-1   & 769.12 & 1.0045 & 1327.15 & 0.0773& -5.583 & 5.77& 1.96\\
-2   & 1160.92& 1.0045 & 1632.50 & 0.07638& -5.762& 5.70& 1.93 \\
-5   & 2363.71& 1.0045 & 2293.39& 0.075451& -5.696& 5.63&  1.91\\
-8   & 3512.61& 1.0045 & 2807.78& 0.075141& -5.674& 5.61& 1.90 \\
-10  & 4296.56& 1.0046 & 3103.70& 0.075028& -5.666& 5.60& 1.89 \\
-15  & 6256.46& 1.0046 & 3742.51& 0.07487& -5.655& 5.59& 1.89 \\
-100 & 39575.31& 1.00457 & 9399.75 & 0.0745753 & -5.633 & 5.56& 1.88 \\
 \hline
\end{tabular}
\end{table}

\begin{table}
\caption{\label{tab5} Model parameters for different anisotropic star. ( $m_0 = 10~M_{\odot}$, $r_0 =80\times1.475~km$ and $k= 0.2$.)} 
\begin{tabular}{cccccccc}
$n$ & $\lambda$ & $R $ & $t_{bh}$ &  $r_{bh}$ & $m_{bh}~(M_{\odot})$\\ \hline
0.80 & 6650.82 & 3203.9 & -4.67102 & 5.80172 &1.967 \\ 
0.50 & 14016.6 & 4622.43 & -4.5475 & 5.6426 & 1.913 \\
0.30 & 18956.6 & 5367.65& -4.51822& 5.60488& 1.899 \\
0.10 & 23924.4& 6024.78& -4.50087 & 5.58252& 1.892 \\
-1 & 51101.1& 8789.3& -4.4653& 5.536& 1.877 \\
-2 & 75810.4& 10699.9& -4.45515& 5.52386& 1.872 \\
-5& 150042& 15044.9& -4.4447& 5.51042&  1.868 \\
-8 & 224251& 18389.6& -4.44115& 5.505& 1.866 \\
-10& 273724& 20315.7& -4.4398& 5.5042& 1.866 \\
-15& 397406& 24476.7& -4.43803& 5.50186& 1.865 \\
 \hline
\end{tabular}
\end{table}

\begin{table}
\caption{\label{tab6} Dynamical variables for different anisotropic star. ( $m_0 =20 ~M_{\odot}$, $r_0 =60\times1.475~km$ and $k = 0.3$.)} 
\begin{tabular}{cccccccc}
$n$ & $\lambda$ & $a$ & $R $ & $h_{bh}$ & $t_{bh}$ &  $r_{bh}$ & $m_{bh}~(M_{\odot})$\\ \hline
   0 & 330.91 & 1.0077 & 1105.80 & 0.1585 & -21.65 & 19.09& 6.47 \\
-1.0 & 661.87 & 1.0079 & 1543.07& 0.1528& -20.95& 18.41& 6.24 \\
-5.0 & 1989.21& 1.0081 &2650.08&0.1488&-20.46&17.93&6.08 \\
-10 & 3649.21&1.0081 &3581.53&0.1479&-20.34&17.82&6.04\\
-20 & 6969.49&1.0081 &4943.37&0.1474&-20.28&17.75&6.02\\
-50 &16930.50&1.0081 &7698.44&0.1470&-20.24&17.71&6.00\\
-100&33532.20&1.0081 &10831.2&0.1469&-20.22&17.70&5.99 \\
\hline
\end{tabular}
\end{table}

\begin{table}[h!]
\centering
\caption{Dynamical variables for different anisotropic star. ($m_0 = 60~M_{\odot}$, $r_0 = 600\times1.475~km$ and $k= 0.1$.)} 
\label{tab7}
\begin{tabular}{cccccccc}
$n$ & $\lambda$ & $a$ & $R $ & $\gamma$ & $t_{bh}$ &  $r_{bh}$ & $m_{bh}~(M_{\odot})$\\ \hline
0.999 & 159 & 0.9999 & 7658 & 0.0000960 & -26.12 & 33.85 & 11.48 \\ 
0.9 & 722  & 1.0004 & 14432 & 0.0000767 & -18.19 & 23.34 & 7.91 \\
0.2 & 4988 & 1.0005 & 36371 & 0.00007005 &  -15.63 & 19.98 & 6.77 \\
-0.1 & 6823 & 1.0005 & 42450 & 0.0000697 &  -15.51 & 19.81 & 6.72 \\
-0.2 & 7435 & 1.0005 & 44292 & 0.0000696 &  -15.48 & 19.78 & 6.70 \\
-0.5 & 9270 & 1.0005 & 49406 & 0.0000695 & -15.41 & 19.69 & 6.68 \\
-0.7 & 10494 & 1.0005 & 52540 & 0.0000694 &  -15.39 & 19.66 & 6.66 \\
-1.0 & 12330 & 1.0005 & 56919 & 0.0000693 &  -15.35 & 19.61 & 6.65 \\
-1.5 & 15389 & 1.0005 & 63550 & 0.0000692 &  -15.31 & 19.56 & 6.63 \\
-2.0 & 18449 & 1.0005 & 69552 & 0.0000691 &  -15.29 & 19.53 & 6.62 \\
-2.5 & 21509 & 1.0005 & 75075 & 0.0000691 & -15.27 & 19.51 & 6.61 \\
-3.0 & 24569 & 1.0005 & 80219 & 0.0000690 & -15.26 & 19.49 & 6.67 \\ \hline
\end{tabular}
\end{table}

\section{Temperature profile}
\label{sec6} 
The laws of thermodynamics during the dynamical gravitational evolutions of physical systems must be obeyed. Black holes are considered to possess remarkable thermodynamic properties. Hawking, Carter and Bardeen by presenting the laws of black hole thermodynamics were successful in establishing a connection between gravity and thermodynamics on the event horizon of a stationary black hole. 

\begin{figure}
\centering
\includegraphics[width=8.0cm, angle=0]{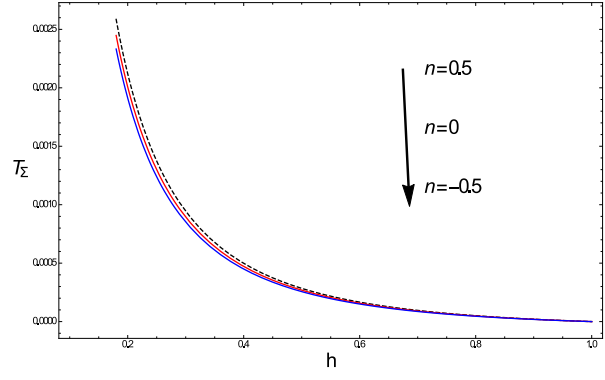}
\caption{Surface temperature profile ($m_0 = 3~M_{\odot}$, $r_0 = 10\times1.475~km$ and $k= 0.2$).}
\label{fg1}
\end{figure}

\begin{figure}
\centering
\includegraphics[width=8.0cm, angle=0]{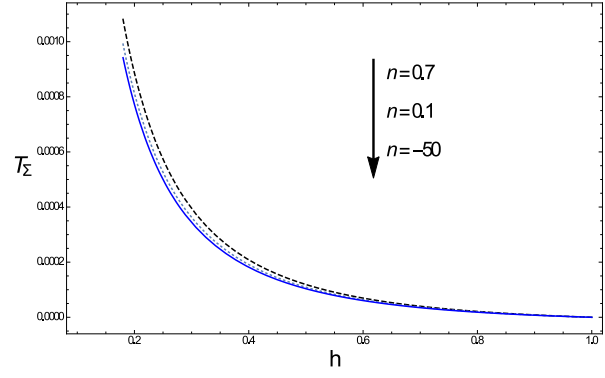}
\caption{Surface temperature profile ($m_0 = 6~M_{\odot}$, $r_0 = 25.742~km$ and $k= 0.2$).}
\label{fg2}
\end{figure}

\begin{figure}
\centering
\includegraphics[width=8.0cm, angle=0]{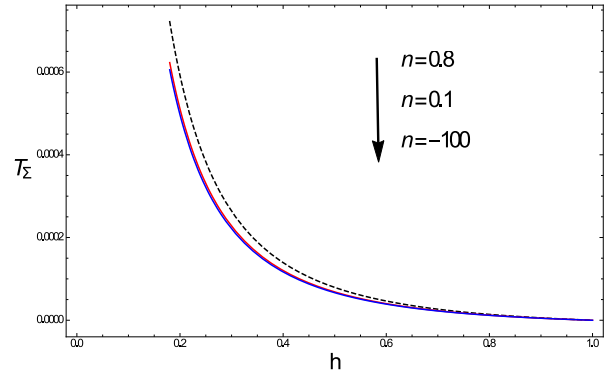}
\caption{Surface temperature profile ($m_0 =10 ~M_{\odot}$, $r_0 =40\times1.475~km$ and $k = 0.2$).}
\label{fg3}
\end{figure}

The temperature evolution of the collapsing system depends on the anisotropy of the system\cite{Sharma+Das+2013}. The relativistic Maxwell-Cattaneo relation for temperature governing the heat transport\cite{Israel+Stewart+1979, Maartens+1995} is given by
\begin{equation}
\tau(g^{\alpha\beta}+u^{\alpha}u^{\beta})u^{\delta}q_{\beta;\delta} + q^{\alpha} = -K(g^{\alpha\beta}+u^{\alpha}u^{\beta})[T_{,\beta}+T\dot{u_{\beta}}],\label{teq41}
\end{equation}
where $T$ is the temperature, $K (\geq 0)$ is the thermal conductivity and $\tau (\geq 0)$ is the relaxation time. For the line element (\ref{intm1}), eq.~(\ref{teq41}) reduces to
\begin{equation}
\tau\frac{d}{dt}(qhB_0) + q h A_0 B_0 = -K \frac{1}{h B_0}\frac{d}{dr}(A_0T).\label{teq42}
\end{equation}
The relativistic Fourier heat transport equation can be simplified by setting $\tau = 0$ in (\ref{teq42}) as shown in Ref.~\cite{Sharma+Das+2013}. For $\tau=0$, combining eqs.~(\ref{fEq4}) and (\ref{teq42}), we get
\begin{equation}
8\pi G K(A_0T)' = \frac{2A_0'\dot{h}}{A_0h}.\label{teq43}
\end{equation}
Let us now assume that the thermal conductivity varies as $K = \beta T^{\omega}$, where $\beta$ and $\omega$ are constants. eq.~(\ref{teq43}), then  yields
\begin{equation}
8\pi\beta(A_0T)'=\frac{2A_0' T^{-\omega}}{A_0}\left[\frac{2\gamma(\sqrt{h}-1)}{h\sqrt{h}}\right],\label{teq44}
\end{equation} 
where we have used eq.~(\ref{surf}). Integrating the above equation, we get
\begin{equation}
T^{\omega+1}=\frac{\gamma(\sqrt{h}-1)}{2\pi\beta h\sqrt{h}}\left(\frac{\ln A_0}{A_0}\right)+T_0(t)
,\label{teq45}
\end{equation} 
To get an overall picture of temperature evolution, we set $\omega=0$, $\beta = 1$ and $T_0(t) = 0$ without loss of any generality. Note that at the onset collapse, in the absence of adequate information about luminosity vis-a-vis thermal profile of the collapsing star, we can set $T_0(t) =0$ so as to analyze the temporal evolution of the system. With the above choices of the  constants the surface temperature at any instant reads as
\begin{equation}
T(r_\Sigma,t)=\frac{\gamma(\sqrt{h}-1)}{2\pi h\sqrt{h}}\left(\frac{\ln A_0}{A_0}\right)_\Sigma
,\label{teq46}
\end{equation} is plotted in Fig.~(\ref{fg1})-(\ref{fg3}). It is evident that the surface temperature which was initially low at $h=1$, gradually increases to  higher values as collapse progresses towards  $h\rightarrow 0$. 

\section{Discussions}
\label{sec7}
The gravitational collapse of a relativistic anisotropic star in the presence of radial heat flux is studied. The star is assumed to begin its collapse from an initial configuration described by Paul and Deb\cite{{pd+et al+2014}} model. In general, it is observed that in the case of anisotropic stars, when the tangential pressure of the initial configuration exceeds the radial pressure ($p_t > p_r$),  {\it i.e.,}  $n > 0$, the horizon is found to form at an earlier epoch compared to that of a less anisotropic star. However, if $p_r > p_t$, {\it i.e.,}  $n < 0$, the horizon is found to form at a later stage for a star with more negative $n$ values. The time of formation of a black hole ($t_{bh}$), the mass of black hole ($m_{bh}$) and corresponding radius ($r_{bh}$) are found to follow similar behaviour for a wide range of massive stars with different compactness factors. 

It is evident from Table~\ref{tab4} and Table~\ref{tab5} that as the compactness factor increases, for a particular mass with a constant $k$, the time of formation of a black hole is found to decrease i.e., it leads to a black hole that forms at a faster rate  as compared to stars with higher compactness.  A star with a larger black hole mass ($m_{bh}$) and size ($r_{bh}$) ends with a high mass. However, for the same $k$ if the anisotropy increases (more negative $n$ values) then it is found that a less massive black hole is formed.

For a given mass and radius, if the geometrical parameter $k$ is increased, it is observed that black hole may form at a faster rate. However, the black hole mass ($m_{bh}$) and the corresponding radius ($r_{bh}$) for such stars ends into a comparately massive black hole for higher values of $k$ for the same anisotropic parameter $n$. 

For a  given mass, as $\lambda$ is increased, the time of formation of a black hole increases. However, the final masses ($m_{bh}$) and radii ($r_{bh}$) of the black holes are found to be identical. It is also found that for a star with an initial configuration having same $k$ and $n$, in the case of a  massive star, the time to form a black hole is less i.e., the black hole is formed at an earlier epoch compared to that of an anisotropic star with lower mass. Accordingly, black hole mass ($m_{bh}$) and black hole radius ($r_{bh}$) both become high for the higher initial massive star.

To conclude, in this paper we have analyzed the effects of anisotropy on the gravitational collapse of a star. The work provides a simple mechanism to investigate the effects of pressure anisotropy for a wide range of initial configurations. The effects of other factors such as shear, viscosity, charge etc. are beyond the scope of the current investigation. We have also studied the anisotropic effect on the temporal behaviour a collapsing system. It is observed that for $n>0$ the surface temperature lies on the higher side as compared to the case of $n < 0$ at all stages of the gravitational collapse.

\begin{acknowledgments}
BCP and RS gratefully acknowledge support from the Inter-University Centre for Astronomy and Astrophysics (IUCAA), Pune, India, under its Visiting Research Associateship Programme.  SD is also thankful to IUCAA, Pune for its hospitality where part of this work was carried out. 
\end{acknowledgments}

\end{document}